\documentclass[a4paper,12pt]{article}
\usepackage{latexsym}
\usepackage{rotating}
\author{Wen-Xiu~Ma\footnote{Email: mawx@math.cityu.edu.hk} \\
Department of Mathematics, City University of Hong Kong,\\
Kowloon, Hong Kong \vspace{2mm} \\
Ruguang Zhou\footnote{Email: rgzhou@public.xz.js.cn}\\
Department  of Mathematics, Xuzhou Normal University  \\ 
Xuzhou 221009, P. R. China }
\title{A Coupled AKNS-Kaup-Newell Soliton Hierarchy}
\date{\nonumber}

\begin{document}
\maketitle

\setlength{\parskip}{5pt plus 2pt minus 1 pt}
  \setlength{\textwidth}{150.5mm}
  \setlength{\textheight}{230mm}
  \setlength{\oddsidemargin}{0mm}
  \setlength{\topmargin}{-25mm}    
\setlength{\baselineskip}{21.5pt}
\date{\nonumber}

\vskip 4mm

\renewcommand{\thesection}{\Roman{section}}
\newtheorem{thm}{Theorem}
\newtheorem{pro}{Proposition}
\newtheorem{Le}{Lemma}
\newtheorem{defi}{Definition}
\newcommand{\R}{\mbox{\rm I \hspace{-0.9em} R}}
\newcommand{\N}{\mbox{\rm I \hspace{-0.8em} N}}

\def\be{\begin{equation}}
\def\ee{\end{equation}}
\def\bea{\begin{eqnarray}}
\def\eea{\end{eqnarray}}
\def\ba{\begin{array}}
\def\ea{\end{array}}
\def\la {\lambda}
\def \part {\partial}
\def \al {\alpha}
\def \de {\delta}


\vskip 5mm

\begin{abstract}
A coupled AKNS-Kaup-Newell hierarchy of systems of soliton equations
is proposed 
in terms of hereditary symmetry operators 
resulted from Hamiltonian pairs.
Zero curvature representations and tri-Hamiltonian structures
are established for all coupled AKNS-Kaup-Newell systems in the hierarchy.
Therefore all systems have infinitely many commuting symmetries
and conservation laws.
Two reductions of the systems lead to the AKNS hierarchy and the Kaup-Newell
hierarchy, and thus 
those two soliton hierarchies also possess
tri-Hamiltonian structures.

\end{abstract}

\section{Introduction}

Systems of soliton equations come usually in hierarchies.
This kind of hierarchies possesses many nice properties, for instance,
Lax representations or zero curvature representations,
infinitely many commuting symmetries and conservation laws,
hereditary recursion structures, bi-Hamiltonian formulations and
even multiple Hamiltonian formulations etc.
and they are often called soliton hierarchies.
Well-known examples of such soliton hierarchies (for example, see
\cite{Newell-book1985,AblowitzC-book1991})
contain the KdV hierarchy, the MKdV hierarchy, the AKNS hierarchy,
the Kaup-Newell hierarchy, the Benjamin-Ono hierarchy \cite{FokasF-PLA1981},
the Tu hierarchy \cite{Tu-SCA1989}, the Dirac hierarchy \cite{Ma-CAMB1997},
the coupled KdV hierarchies \cite{BoitiPC-NCB1983,AntonowiczF-PD1987},
the coupled Harry-Dym hierarchies \cite{AntonowiczF-JPA1988},
the coupled Burgers hierarchies \cite{Ma-JPA1993} and so on.
It is very interesting to search for new soliton hierarchies,
even hierarchies of systems
which possess only infinitely many commuting symmetries.

An idea which allows to achieve this is to construct soliton hierarchies
of coupled systems of equations. 
It could be divided into two aspects in view of types of soliton equations. 
The one is to construct soliton hierarchies by 
coupling systems of the same type.
Such examples are the coupled KdV hierarchies
\cite{BoitiPC-NCB1983,AntonowiczF-PD1987}, the coupled Harry-Dym hierarchies 
\cite{AntonowiczF-JPA1988},
the coupled Burgers hierarchies
\cite{Ma-JPA1993}, and the perturbation systems of the KdV hierarchy
\cite{MaF-PLA1996} etc.
The other is to construct soliton hierarchies by 
coupling systems of different types. There are few examples in this 
aspect. A coupled AKNS-Kaup-Newell
hierarchy of complex form, recently introduced by Zhang
\cite{Zhang-ADE1997}, is such an example.

In this paper, motivated by Zhang's coupled AKNS-Kaup-Newell hierarchy
of complex form,
we would like to propose a hierarchy of
coupled AKNS-Kaup-Newell evolution equations
of real form.
The hierarchy will be established in the second section,
in terms of hereditary symmetry operators. The required
hereditary symmetry operators can be generated by
observing a set of Hamiltonian operators.
Zero curvature representations will be computed in the third section 
for all systems in the hierarchy. 
Interestingly the discussion of the fourth section
shows that all the systems have not only bi-Hamiltonian 
structures but also tri-Hamiltonian structures,
although Zhang didn't present Hamiltonian structures and consequent 
conservation laws due to a failure in determining
Hamiltonian operators \cite{Zhang-ADE1997}. 
Therefore the resulting hierarchy has infinitely
many commuting symmetries and
conservation laws.
Some concluding remarks
are given in the last section. 

\section{Hereditary symmetry operators}

\label{Hamiltonianoperators}

We want to establish a coupled AKNS-Kaup-Newell hierarchy 
in terms of hereditary symmetry operators resulted from Hamiltonian pairs.
To this end, let us introduce
a set of $2\times 2$ matrix integro-differential operators:
\be    {M} =M(u)= \left (\ba {cc}  \al _1 q\part ^{-1}q & \al _2+\al _3\part
-\al _1q\part ^{-1}r \vspace{2mm}\\
 -\al _2+\al _3 \part -\al _1 r\part ^{-1}q & \al _1 r\part ^{-1}r 
 \ea  \right),\
 u=  \left (\ba {cc} q  \vspace{2mm}\\ r \ea \right ),\label{bigoperator}
  \ee
where $ \part  =\frac{\part }{\part x},\,q=q(x,t),\, r=r(x,t),$ and 
$\al _1,\al _2,\al _3$ are three arbitrary constants,
and consider their Hamiltonian property.
They are simple generalizations of the Hamiltonian operators in the AKNS case
\cite{Magri-book1980}. 
The following proposition shows that they are still Hamiltonian, indeed.

\begin{pro} \label{HamiltonianoperatorofgeneralAKNS}
The $2\times 2$ matrix integro-differential operators defined by
(\ref{bigoperator}) 
are all Hamiltonian for any constants $\al _1,\al _2,\al _3$.
\end{pro}
{\bf Proof:}
Assume that
 \[X=(X_1,X_2)^T,\ Y=(Y_1,Y_2)^T,\ Z=(Z_1,Z_2)^T,\ W=(W_1,W_2)^T, \]
are two-dimensional vectors of functions.
Since $M$ is skew-symmetric, we only need to prove that the Jacobi identity 
\be <Z,M'[MX]Y>+\rm{cycle}(X,Y,Z)\equiv 0\ (\rm{mod}\, \part ) 
\label{jacobiidentity}\ee
holds for any $X,Y,Z$, where $<\cdot,\cdot>$ denotes the standard inner product
of $\R ^2$.
By (\ref{bigoperator}), we immediately have
\[  MX=\left (\ba {c} \al _2X_2+\al _3X_{2x}+\al _1qP(X)\vspace{2mm}\\ 
-\al _2X_1+\al _3X_{1x}-\al _1rP(X)
\ea \right ) :=\left (\ba {c} W_1(X)\vspace{2mm}\\ W_2(X)
\ea \right) , \]
where $P(X)=\part ^{-1}(qX_1-rX_2)$. Following the definition of the Gateaux
derivative, two objects $M'[W]$ and $M'[W]Y$ are computed as follows: 
\[\ba {l} M'[W] = \left(\ba {cc} \al _1q\part ^{-1}W_1+\al _1W_1\part ^{-1}q
&  - \al _1q\part ^{-1}W_2-\al _1W_1\part ^{-1}r
\vspace{2mm}\\
-\al _1r\part ^{-1}W_1-\al _1W_2\part ^{-1}q
&  \al _1r\part ^{-1}W_2+\al _1W_2\part ^{-1}r
\ea \right ),\vspace{2mm}\\
 M'[W]Y = \left(\ba {c} \al _1q\part ^{-1}(W_1Y_1-W_2Y_2)+
 \al _1W_1\part ^{-1}(qY_1-rY_2)
\vspace{2mm}\\
-\al _1r\part ^{-1}(W_1Y_1-W_2Y_2)-\al _1W_2\part ^{-1}(qY_1-rY_2)
\ea \right ).
\ea \]
Now we can have
\bea 
<Z,M'[MX]Y>&=& \al _1(qZ_1-rZ_2)\part ^{-1}(W_1(X)Y_1-W_2(X)Y_2)
\nonumber \\
&& +\al _1(W_1(X)Z_1-W_2(X)Z_2)\part ^{-1}(qY_1-rY_2). \eea
Upon observing that 
\bea && W_1(X)Y_1-W_2(X)Y_2 \nonumber \\
&=& (\al _2X_2+\al _3X_{2x}+\al _1qP(X))Y_1
-(-\al _2X_1+\al _3X_{1x}-\al _1rP(X))Y_2
\nonumber \\
&=& \al _1P(X)(qY_1+rY_2)+\al _2(X_2Y_1+X_1Y_2)+\al _3(X_{2x}Y_1-X_{1x}Y_2),
\nonumber
\eea
 we can make a decomposition
\be <Z,M'[MX]Y>= R(X,Y,Z)+  S(X,Y,Z)+  T(X,Y,Z), 
\label{decompositionfortotal}\ee
where $R,S,T$ are defined by
\bea
 R(X,Y,Z)&=& \al _1^2(qZ_1-rZ_2)\part ^{-1} \bigl [P(X)(qY_1+rY_2) \bigr]
\nonumber\\
&& +\al _1^2P(X)(qZ_1+rZ_2)\part ^{-1} (qY_1-rY_2),
\\
S(X,Y,Z)&=&
\al _1\al _2(qZ_1-rZ_2)\part ^{-1} (X_2Y_1+X_1Y_2)
\nonumber\\
&& +\al _1\al _2(X_2Z_1+X_1Z_2)\part ^{-1} (qY_1-rY_2),
\\
 T(X,Y,Z)&=& \al _1\al _3(qZ_1-rZ_2)\part ^{-1} (X_{2x}Y_1-X_{1x}Y_2)
\nonumber\\
&& +\al _1\al _3(X_{2x}Z_1-X_{1x}Z_2)\part ^{-1} (qY_1-rY_2).
\eea
For these three functions $R,S,T$, we can compute that
\bea &&
R(X,Y,Z)+\rm{cycle}(X,Y,Z)
\nonumber \\
&=& \al _1^2\part \{ P(Z)\part ^{-1}\bigl[ P(X)(
qY_1+rY_2)\bigr]\}
+\rm{cycle}(X,Y,Z),\label{Rpart}\vspace{1mm} \\
&&S(X,Y,Z)+\rm{cycle}(X,Y,Z)
\nonumber \\
&=& \al _1\al _2\part \bigl[ P(Z)\part ^{-1}(X_2Y_1+X_1Y_2)\bigr]
+\rm{cycle}(X,Y,Z),\label{Spart} \vspace{1mm}\\
&& T(X,Y,Z)+\rm{cycle}(X,Y,Z)
\nonumber \\
&=& \al _1\al _3(qZ_1-rZ_2)(X_2Y_1-X_1Y_2)
+  \al _1\al _3(qZ_1-rZ_2)\part ^{-1}(X_1Y_{2x}-X_2Y_{1x})\nonumber \\
&&+ \al _1\al _3(Z_1X_{2x}-Z_2X_{1x})\part ^{-1}(qY_1-rY_2)
+\rm{cycle}(X,Y,Z)
\nonumber \\
&=&  \al _1\al _3(qZ_1-rZ_2)(X_1Y_{2x}-X_2Y_{1x})\nonumber \\
&&+ \al _1\al _3(Z_1X_{2x}-Z_2X_{1x})\part ^{-1}(qY_1-rY_2)
+\rm{cycle}(X,Y,Z)
\nonumber \\
&=&  \al _1\al _3\part \bigl[ P(Z)\part ^{-1}(X_1Y_{2x}-X_2Y_{1x})\bigr]
+\rm{cycle}(X,Y,Z).
\label{Tpart} 
\eea
They are all total derivatives and thus 
combining the decomposition (\ref{decompositionfortotal}) and  the equalities
(\ref{Rpart}), (\ref{Spart}), (\ref{Tpart}) 
leads to the Jacobi identity (\ref{jacobiidentity}).
This completes the proof.
$\vrule width 1mm height 3mm depth 0mm$

Now we would like 
to discuss some special cases of Hamiltonian pairs starting from
the above Hamiltonian operators defined by (\ref{bigoperator}),
which allows to generate 
hereditary symmetry operators and further soliton hierarchies.  
This idea has been successfully 
used to construct bi-Hamiltonian coupled KdV systems
\cite{Ma-1998,MaP-1998}.
An important phenomenon we want to point out 
is that different soliton hierarchies can be derived 
from Hamiltonian operators of the same type. 
The following discussion shows an example of such phenomenon. 
It is also important to realize that not all Hamiltonian pairs 
may generate hereditary symmetry operators. 
Thus care must be taken to restrict our 
attention to the cases where there exists at least one
invertible Hamiltonian operator
for each Hamiltonian pair.
The required invertibility guarantees 
that Hamiltonian pairs can yield hereditary
symmetry operators \cite{FuchssteinerF-PD1981}.
  
{\bf Case 1:}
We make a choice of an invertible Hamiltonian operator
\be
 {J} = \left (\ba {cc} 0&1 \vspace{2mm}\\ -1 &0 
 \ea  \right),\ee
which has an inverse operator
\[
 {J^{-1}} = \left (\ba {cc} 0&-1 \vspace{2mm}\\ 1 &0 
 \ea  \right).
   \]
It follows from Proposition \ref{HamiltonianoperatorofgeneralAKNS} 
that this operator $J$
constitutes a Hamiltonian pair
with $M$ defined by (\ref{bigoperator}). 
Therefore we can have a hereditary symmetry operator
\be
 \Phi  = MJ^{-1}=\left (\ba {cc} \al _2+\al _3\part -\al _1q\part ^{-1}
 r &-\al _1 q \part ^{-1}q  \vspace{2mm}\\ \al _1r \part ^{-1}r  &
 \al _2-\al _3\part +\al _1r\part ^{-1}q
 \ea  \right),\ee
where $\al _1,\al _2$ and $\al _3$ are arbitrary.
The reduction of $\al _1=-1$, $\al _2=0$ and $\al _3=\frac 12 $
leads to the recursion operator for the normal AKNS hierarchy
\cite{Newell-book1985,AblowitzC-book1991,MaS-PLA1994}.

{\bf Case 2:}
We make a choice of a Hamiltonian pair
\be
 {J} = \left (\ba {cc} \beta _1 q\part ^{-1}q &1-\beta _1
q \part ^{-1}r \vspace{2mm}\\ -1-\beta _1
r \part ^{-1}q   & \beta _1 r\part ^{-1}r  
 \ea  \right),\ 
 {M} = \left (\ba {cc} 0&\al _3 \part  \vspace{2mm}\\ 
\al _3\part &0 
 \ea  \right).
\ee
The above proposition ensures 
that they constitute a Hamiltonian pair, indeed.
Since the operator $J$ has an invertible operator
\[
 {J^{-1}} = \left (\ba {cc}  \beta _1 r\part ^{-1}r &
-1+\beta _1r\part ^{-1}q \vspace{2mm}\\ 1+\beta _1q\part ^{-1}r &
\beta _1q\part ^{-1}q 
 \ea  \right),
   \]
we can obtain the corresponding hereditary symmetry operator
\be
 \Phi  = MJ^{-1}=\left (\ba {cc} \al _3\part +\al _3\beta _1
\part q\part ^{-1}r&
  \al _3\beta _1\part  q \part ^{-1}q  \vspace{2mm}\\
 \al _3\beta _1\part r \part ^{-1}r  & -\al _3\part +\al _3
 \beta _1r\part ^{-1}r
 \ea  \right),\ee
where $\al _3$ and $\beta _1$ are arbitrary.
The reduction of $\al _3=\frac12 $ and $\beta_1 =-1$
leads to the recursion operator for the normal Kaup-Newell hierarchy
(for example, see \cite{MaDZL-NCB1996}).

More generally, we have the following case, which combines the above two cases.
 
{\bf Case 3:}
We make a choice of an invertible Hamiltonian operator
\be
 {J} = \left (\ba {cc} \beta _1 q\part ^{-1} q&
 \beta _2 -\beta _1 q\part ^{-1}r  \vspace{2mm}\\
 -\beta _2-\beta _1r\part^{-1} q &\beta _1r\part ^{-1}r
 \ea  \right),\ee 
whose inverse operator can be given by
\[ 
 {J^{-1}} = \frac 1{\beta _2^2}\left (\ba {cc}
 \beta _1 r\part ^{-1} r&
- \beta _2 +\beta _1 r\part ^{-1}q  \vspace{2mm}\\
\beta _2+\beta _1q\part^{-1} r &\beta _1q\part ^{-1}q
 \ea  \right).
   \]
It follows from Proposition
\ref{HamiltonianoperatorofgeneralAKNS} that the operator $J$
constitutes a Hamiltonian pair
with the Hamiltonian operator $M$ defined by (\ref{bigoperator}). 
In this case, we can generate
the following corresponding hereditary symmetry operator
\bea
 & \Phi  = MJ^{-1}=\frac 1 {\beta _2^2}&
 \left (\ba {c} \al _2\beta _2 +\al _3\beta _2\part +(\al _2\beta _1-
 \al _1\beta _2)q\part ^{-1}r+\al _3\beta _1\part q\part ^{-1}r
, \vspace{2mm}\\
 (\al _1\beta _2- \al _2\beta _1)r\part ^{-1}r +\al _3\beta _1\part r
 \part ^{-1}r , \ea \right .
 \nonumber \\  &&
\  \left. \ba {c}
(\al _2\beta _1-
 \al _1\beta _2)q\part ^{-1}q+\al _3\beta _1\part q\part ^{-1}q
 \vspace{2mm}\\
 \al _2\beta _2 -\al _3\beta _2\part +(\al _1\beta _2-
 \al _2\beta _1)r\part ^{-1}q+\al _3\beta _1\part r\part ^{-1}q
 \ea  \right), \quad \eea
where five constants are arbitrary but $\beta _2\ne 0$.

Let us pick out 
a special sub-case of $\al _2=0$ and $ \beta _2=1$ from the third case.
If $\al _3=0$, we just obtain a simple hereditary symmetry operator
\be  \Phi =\left (\ba {cc} -\al  q\part ^{-1}r& -\al  q\part ^{-1}q
\vspace {2mm}\\
  \al  r\part ^{-1}r& \al  r\part ^{-1}q
\ea \right ), \label{specialrecursionoperatorAKNS}\ee
where $\al =\al _1$ is arbitrary. This is equivalent to the first case 
with $\al _2=\al _3=0$.

If $\al _3\ne 0$, upon resetting $\al _3=\gamma ,\,\al _1=\al ,\,
\al _3\beta _1=\beta $, we obtain a hereditary symmetry operator
\be  \Phi =\left (\ba {cc} \gamma \part -\al  q\part ^{-1}r
+\beta \part q\part ^{-1}r&
-\al  q\part ^{-1}q +\beta \part q\part ^{-1}q \vspace {2mm}\\
  \al  r\part ^{-1}r+\beta \part r\part ^{-1}r& -\gamma \part
  +\al  r\part ^{-1}q+\beta \part r\part ^{-1} q
\ea \right ), \label{hereditaryoperator3}\ee
where $\al ,\beta,\gamma$ are arbitrary but $\gamma \ne 0$.
Note that if we let the constant $\gamma $ go to zero,
the hereditary condition for $\Phi $ with a general constant $\gamma $
becomes the one for $\Phi $ with $\gamma =0$.
Therefore the constant $\gamma $ can be chosen as zero,
which doesn't affect the hereditary property of $\Phi$. 
However if $\gamma =0$, we don't know whether the operator $\Phi $ defined by 
(\ref{hereditaryoperator3}) is decomposable, i.e.
whether there exists any Hamiltonian pair $J$ and $M$
so that $\Phi =MJ^{-1}$.

We will focus on discussing a soliton hierarchy generated by 
the hereditary symmetry operator in (\ref{hereditaryoperator3})
because of its generality.
For $\gamma \ne 0$, we can re-scale three constants 
to put a general case into a special case of the operator $\Phi $ defined by
(\ref{hereditaryoperator3}), and thus we
pick out the following special case 
\be  
 \Phi =MJ^{-1}=\left (\ba {cc}
\frac 12 \part -\al q\part ^{-1}r - \frac 12 \beta \part q\part ^{-1}r &
- \al q\part ^{-1} q- \frac 12 \beta \part q \part ^{-1}q
\vspace{2mm}\\
 \al r\part ^{-1} r -  \frac 12 \beta \part r\part ^{-1}r &
- \frac 12 \part + \al r\part ^{-1} q- \frac 12 \beta \part r \part ^{-1}q
 \ea  \right)
\label{hereditaryoperatorAKNSKN}\ee
to discuss 
without loss of generality.
To this special case, the corresponding hierarchy of evolution equations 
\be 
\left(\ba {c}q_t\vspace{2mm}\\ r_t\ea \right )=
\left (\ba {cc}
\frac 12 \part -\al q\part ^{-1}r - \frac 12 \beta \part q\part ^{-1}r &
- \al q\part ^{-1} q- \frac 12 \beta \part q \part ^{-1}q
\vspace{2mm}\\
 \al r\part ^{-1} r -  \frac 12 \beta \part r\part ^{-1}r &
- \frac 12 \part + \al r\part ^{-1} q- \frac 12 \beta \part r \part ^{-1}q
 \ea  \right)^n
\left(\ba {c}q_x\vspace{2mm}\\ r_x\ea \right ),\ n\ge 0,
\label{hierarchyAKNSKN} \ee
contains two important reductions.
If $\al \ne0$ but $\beta = 0$, the hierarchy reduces to the AKNS hierarchy. 
If $\al =0$ but $\beta \ne 0$, the hierarchy reduces to
the Kaup-Newell hierarchy. Thus the hierarchy 
(\ref{hierarchyAKNSKN}) generated by the hereditary symmetry operator
(\ref{hereditaryoperatorAKNSKN})
is called a coupled AKNS-Kaup-Newell hierarchy.
All systems in the hierarchy (\ref{hierarchyAKNSKN})
are real. Therefore the hierarchy (\ref{hierarchyAKNSKN})
is a soliton hierarchy that we want to construct.

\section{Zero curvature representations}

\label{ZCRofcoupledAKNSKN}
In the previous section, we generated a coupled AKNS-Kaup-Newell hierarchy
of real form by observing Hamiltonian operators. More importantly, 
the resulting hierarchy shares some common integrable
properties. In this section, we want to show zero curvature 
representations for all systems in the hierarchy, and in the next section,
we will establish tri-Hamiltonian structures.

To show zero curvature representations,
let us impose a spectral problem
\be \phi _x=U\phi,
\ U=U(u,\la )=\left(
\ba {cc} \la & q \vspace{2mm}\\ (\al +\beta \la )r & -\la 
\ea \right), \ \phi=
\left(\ba {c} \phi _1\vspace{2mm}\\ \phi _2
\ea \right), \ee
where $\la $ is a spectral parameter, and $\al $ and $\beta $
are arbitrary constants.
It is customary to 
solve the stationery zero curvature equation
$ V_x=[U,V]$ first.
Suppose that
\be V=V(u,\la )=\left(\ba {cc} a & b \vspace{2mm}\\ (\al +\beta \la )c & -a
\ea \right)=\sum_{i\ge 0}\left(
\ba {cc} a_i & b_i \vspace{2mm}\\ (\al +\beta \la )
c_i & -a_i \ea \right )\la ^{-i},
\ee
and then the stationery zero curvature equation
becomes
\be \left\{ \ba {l} a_x=(\al +\beta \la )(qc-rb), \vspace{2mm} \\
 b_x=2\la  b-2qa, \vspace{2mm} \\
  c_x=2ra-2\la c.
\ea \right. \label{recursionrelationforabc}\ee
Notice that a recursion relation to determine $b$ and $c$ may be found 
if we fix $a=(\al +\beta \la )\part ^{-1}(qc-rb)$.
Actually we have
\be  \left \{ \ba {l} b_x= 2\la b -2 (\al +\beta \la )q\part ^{-1}
(qc-rb),
 \vspace{2mm}\\
c_x= 2 (\al +\beta \la )r\part ^{-1}
(qc-rb)-2\la c ,
 \ea \right. \nonumber \ee
which equivalently leads to
\be \left(\ba {cc} -2\beta q\part ^{-1}q & 2+2\beta q\part ^{-1}r
 \vspace{2mm}\\ -2+2\beta r\part ^{-1}q & -2\beta r \part ^{-1}r \ea \right)
\left( \ba {c}  c_{i+1} \vspace{2mm}\\ b_{i+1} \ea \right) =
\left(\ba {cc} 2\al  q\part ^{-1}q & \part -2\al  q\part ^{-1}r
 \vspace{2mm}\\ \part -2\al  r\part ^{-1}q & 2\al  r \part ^{-1}r \ea \right)
 \left(\ba {c}  c_{i} \vspace{2mm}\\ b_{i} \ea \right), \ee
where $i\ge 0$.
If we set
\be J=\left(\ba {cc} -2\beta q\part ^{-1}q & 2+2\beta q\part ^{-1}r
 \vspace{2mm}\\ -2+2\beta r\part ^{-1}q & -2\beta r \part ^{-1}r \ea
\right),\ M=
\left (\ba {cc} 2\al  q\part ^{-1}q & \part -2\al  q\part ^{-1}r
 \vspace{2mm}\\ \part -2\al  r\part ^{-1}q & 2\al  r \part ^{-1}r \ea
\right),\label{HamiltonianpairAKNSKN}\ee
the operators $J$ and $M$ constitute
a Hamiltonian pair, based on the result in the previous section.
It is apparent that the corresponding hereditary symmetry operator
$\Phi =MJ^{-1}$ is exactly the same as the one defined by 
(\ref{hereditaryoperatorAKNSKN}),
having noted that 
\[ J^{-1}= \frac 12 \left(
\ba {cc} - \beta \part r\part ^{-1}r &-1 - \beta \part r\part ^{-1}q
\vspace{2mm}\\
1- \beta \part q\part ^{-1}r& - \beta \part q\part ^{-1}q 
\ea \right) .\]
The conjugate operator of $\Phi $ reads as 
\be \Psi =\Phi ^\dagger =J^{-1}M=\left( \ba {cc}
-\frac 12 \part +\al r\part ^{-1}q - \frac 12 \beta r\part ^{-1}q\part  &
- \al r\part ^{-1} r-\frac 12\beta r \part ^{-1}r\part 
\vspace{2mm}\\
 \al q\part ^{-1} q - \frac 12 \beta q\part ^{-1}q\part   &
\frac 12 \part - \al q\part ^{-1} r-\frac 12 \beta q \part ^{-1}r\part 
 \ea \right). \ee
Therefore upon noting (\ref{recursionrelationforabc}) and choosing $a_0=1$,
we obtain a solution to the stationary zero curvature equation $V_x=[U,V]$:
\be \left \{ \ba {l}
a_0=1,\ b_0=c_0=0;  \ b_1=q,\ c_1=r;
\vspace{2mm}\\
\left (\ba {c} c_{i+1} \vspace{2mm}\\ b_{i+1} \ea \right)=\Psi
\left (\ba {c} c_{i} \vspace{2mm}\\ b_{i} \ea \right),\
i\ge 1,
\vspace{2mm}\\
a_{i}=\al \part ^{-1}(qc_i-rb_i)+\beta \part ^{-1}(qc_{i+1}-rb_{i+1}),
\ i\ge 1;
\ea   \right.   \ee
from which we can get
\[ 
\left (\ba {c} c_2 \vspace{2mm}\\ b_2 \ea \right)=\Psi
\left (\ba {c} r \vspace{2mm}\\ q \ea \right)
=\frac 12 \left (\ba {c} -r_x-\beta qr^2 \vspace{2mm}\\
q_x-\beta q^2r  \ea \right) ,
\]
and 
\[
a_1=\beta \part ^{-1} (qc_2-rb_2)=
-\frac 12  \beta qr.
 \]
It should be noted that we always need to
select zero constants for integration in deriving $a_i,b_i,c_i,\,i\ge 1.$
That is, we require that $a_i|_{[u]=0}=b_i|_{[u]=0}=c_i|_{[u]=0}=0,\ i\ge 1,$
where $[u]=(u,u_x,\cdots)$.

Now we can express the coupled AKNS-Kaup-Newell hierarchy 
(\ref{hierarchyAKNSKN}) in another way.
Let us define
\be  u_t=K_n:= J\left (\ba {c}c_{n+1}\vspace{2mm}\\ b_{n+1} \ea \right)=
\Phi ^n \left(\ba {c}2q \vspace{2mm}\\ -2r\ea \right)
 ,\ n\ge 0, \label{generalhierarchyAKNSKN}\ee
where $\Phi $ is defined by (\ref{hereditaryoperatorAKNSKN}).
The first three systems of the hierarchy (\ref{generalhierarchyAKNSKN}) are
\be \left \{\ba {l}
u_t=
\left( \ba {c} q_t\vspace{2mm}\\ r_t \ea \right)=
K_0= J\left (\ba {c}c_1\vspace{2mm}\\ b_1 \ea \right)=
\left (\ba {c}2q\vspace{2mm}\\ -2r \ea \right) ,
\vspace{2mm}\\
u_t=
\left( \ba {c} q_t\vspace{2mm}\\ r_t \ea \right)=
K_1=
J\left (\ba {c}c_2\vspace{2mm}\\ b_2 \ea \right)=
\left (\ba {c}q_x\vspace{2mm}\\ r_x \ea \right),
\vspace{2mm}\\
u_t
=\left( \ba {c} q_t\vspace{2mm}\\ r_t \ea \right)
=K_2=
J\left (\ba {c}c_3\vspace{2mm}\\ b_3 \ea \right)=
\frac1 2\left (\ba {c} q_{xx}-2\al q^2r-\beta (q^2r)_x
\vspace{2mm}\\  -r_{xx}+2\al qr^2-\beta (qr^2)_x \ea \right).
\ea \right.\ee
Since $K_1=u_x$, all systems in 
the hierarchy (\ref{generalhierarchyAKNSKN}), except the first system
$u_t=K_0$,
are exactly the coupled AKNS-Kaup-Newell  systems in the hierarchy 
(\ref{hierarchyAKNSKN}). Therefore (\ref{generalhierarchyAKNSKN})
is another expression for the 
coupled AKNS-Kaup-Newell hierarchy 
(\ref{hierarchyAKNSKN}).

Let us turn to construction of  
zero curvature representations
for all coupled AKNS-Kaup-Newell systems in the soliton hierarchy 
(\ref{generalhierarchyAKNSKN}).
We need a condition of $\al ^2+\beta ^2\ne 0$. With this condition,
we have the injective property of the Gateaux derivative of $U$ with
respect to $u$, which is required in deriving systems of evolution equations
from zero curvature equations. 
If the condition of $\al ^2+\beta ^2\ne 0$ is not satisfied,
then the systems defined by
(\ref{generalhierarchyAKNSKN}) are  linear and separated, and thus they
are all trivial.

We choose Lax operators $V^{(n)}$ for $n\ge 0$ as
\bea &&
V^{(n)}=  V^{(n)}(u,\la )=\bar V ^{(n)}+\Delta_n ,\
\Delta_n =
 \left (\ba {cc} \delta _{1n} & 0 \vspace{2mm}\\ 0
& \delta _{2n} \ea \right),
\\ &&
\bar V ^{(n)}=
\sum_{j=0}^n \left (\ba {cc} a_j&b_j \vspace{2mm}\\ (\al +\beta \la )c_j
& -a_j \ea \right)\la ^{n-j}
= \left (\ba {cc} (\la ^n a)_+&(\la ^nb)_+ \vspace{2mm}\\
(\al +\beta \la )(\la ^n c)_+
& -(\la ^n a)_+ \ea \right), \quad
\eea
where the subscript denotes to choose the polynomial part in
$\la $, and $\delta_{1n}$ and $\delta_{2n}$ are two functions
to be determined. At this moment, we can compute that
\[ \ba {l}
\bar V ^{(n)}_x-[U, \bar V ^{(n)}]
= \left (\ba {cc} a_{nx}- \al (qc_n -rb_n) &
b_{nx}+2qa_n \vspace{2mm}\\
(\al +\beta \la )( c_{nx}-2ra_n)
& - a_{nx}+\al (qc_n-rb_n) \ea \right), 
\vspace{2mm}\\
\Delta_{nx}-\bigl[ U,\Delta_n  \bigr]
= \left (\ba {cc} \delta _{1n,x} & q(\delta _{1n}- \delta _{2n})
\vspace{2mm}\\ -
(\al +\beta \la )  r(\delta _{1n}- \delta _{2n})
&\delta _{2n,x} \ea \right).
\ea \]
Therefore if we take a choice
\be \delta _{1n}=-\delta _{2n}= -a_n+\al \part ^{-1}(qc_n-rb_n),\ n\ge 1,
\ee
then noting the injective property of $U'$
under $\al ^2+\beta ^2\ne 0$,
the zero curvature equation
\be U_t-V^{(n)}_x+[U,V^{(n)}]=0,\label{zcrofAKNS-KAUP-Newell} \ee
equivalently yields
the coupled AKNS-Kaup-Newell system
\[  u_t=K_n=M\left(\ba {c} c_n\vspace{2mm} \\ b_n\ea
 \right)=J\left(\ba {c} c_{n+1}\vspace{2mm} \\ b_{n+1}\ea
 \right)\]
for each $n\ge 1$.
Moreover it is easy to see that $u_t=K_0$ has a Lax pair
$U$ and $\bar V^{(0)}$. Therefore 
each coupled AKNS-Kaup-Newell system
$u_t=K_n$ has a zero curvature representation with the Lax pair 
$U$ and $V^{(n)}$ if we adopt $\delta _{10}=\delta _{20}=0$.
We remark that 
for the systems $u_t=K_n$ with
$\al =\beta =0,\, n\ge 0,$ 
the above zero curvature representations
still hold, but they are not 
sufficient, because we lose the injective property of
$U'$ in the case of $\al =\beta =0$.

\section{Tri-Hamiltonian structures}

To establish some kind of tri-Hamiltonian structures
for the coupled AKNS-Kaup-Newell hierarchy,
let us impose a third Hamiltonian operator
\bea &
N=M\Psi = &
\left ( \ba {c} -\al q\part ^{-1}q\part +\al \part q \part ^{-1}q
-\frac 12 \beta \part q\part ^{-1} q\part ,
 \vspace{2mm} \\  -\frac 12 
\part ^2 +\al \part r\part ^{-1}q +\al r\par ^{-1}
q\part -\frac 12 \beta \part r \part ^{-1}q\part ,
\ea \right . \vspace{1mm} \nonumber \\
&&
\left . \ba {c} \frac 12 
 \part ^2 -\al  q\part ^{-1}r\part  -\al \part q\part ^{-1}r
  -\frac 12 \beta \part q \part ^{-1}r\part \vspace{2mm}\\
-\al \part r\part ^{-1}r +\al r \part ^{-1}r\part 
-\frac 12 \beta \part r\part ^{-1} r\part 
\ea \right ) . \eea
It constitutes a Hamiltonian triple with $J$ and $M$ defined by 
(\ref{HamiltonianpairAKNSKN}), for any constants $\al ,\beta ,
\gamma $.
That is,
any linear combination of $J$, $M$ and $N$ is still a Hamiltonian operator,
which is automatically satisfied since $J$ and $M$ are a Hamiltonian pair.

Let us consider the first nonlinear system in the
couple AKNS-Kaup-Newell hierarchy (\ref{hierarchyAKNSKN}):
\be
u_t=K_2=M
\left (\ba {c} c_2\vspace{2mm}\\ b_2\ea \right)=
\frac1 2\left (\ba {c} q_{xx}-2\al q^2r-\beta (q^2r)_x
\vspace{2mm}\\  -r_{xx}+2\al qr^2-\beta (qr^2)_x \ea \right).
\label{K_2ofcoupledAKNS-Kaup-Newellhierarchy}\ee
It is apparent that this system could be written in three ways as
\[u_t= K_2=J
\left (\ba {c} c_3\vspace{2mm}\\ b_3\ea \right) =
M\left (\ba {c} c_2\vspace{2mm}\\ b_2\ea \right)
=N\left (\ba {c} c_1\vspace{2mm}\\ b_1\ea \right).
 \]
Moreover a direct calculation can show three gradient vectors 
\bea
&&
\left (\ba {c} c_1\vspace{2mm}\\ b_1\ea \right)=
\left (\ba {c} r\vspace{2mm}\\ q\ea \right)=
\frac {\delta H_0}{\delta u},\ H_0=qr;
\label{H_0AKNSKN} \\ &&
\left (\ba {c} c_2\vspace{2mm}\\ b_2\ea \right)=\Psi
\left (\ba {c} c_1\vspace{2mm}\\ b_1\ea \right)=\frac 12 
\left (\ba {c} -r_{x}-\beta qr^2\vspace{2mm}\\ q_x-\beta q^2r\ea \right)=
\frac {\delta H_1}{\delta u},\nonumber \\
&&\quad  H_1=-\frac 14 \beta q^2r^2-\frac 14 qr_x+\frac 14 q_xr;
\label{H_1AKNSKN}\\ &&
\left (\ba {c} c_3\vspace{2mm}\\ b_3\ea \right)=\Psi
\left (\ba {c} c_2\vspace{2mm}\\ b_2\ea \right)=\frac 14 
\left (\ba {c} r_{xx}-2\al qr^2+3\beta qrr_x+\frac 32 \beta ^2 q^2r^3
\vspace{2mm}\\ q_{xx}-2\al q^2r-3\beta qq_xr +\frac 32 \beta ^2q^3r^2
\ea \right)=
\frac {\delta H_2}{\delta u},\quad \nonumber \\
&& \quad H_2=\frac 18 qr_{xx}+\frac 18 q_{xx}r -\frac 14 \al q^2r^2
+\frac 3 {16}\beta q^2rr_x -
\frac 3 {16}\beta qq_xr^2+\frac 18 \beta ^2 q^3r^3. 
\label{H_2AKNSKN}\eea
Therefore a tri-Hamiltonian structure for the coupled AKNS-Kaup-Newell
system (\ref{K_2ofcoupledAKNS-Kaup-Newellhierarchy})
can be given by
\be
u_t=K_2=
J\frac {\delta H_2}{\delta u}= M  \frac {\delta H_1}{\delta u}
=N\frac {\delta H_0}{\delta u} ,
 \ee
where three Hamiltonian functions $H_0$\, $H_1$ and $H_2$ are defined by 
(\ref{H_0AKNSKN}), (\ref{H_1AKNSKN}) and (\ref{H_2AKNSKN}), respectively.
Based on the recursion scheme in
\cite{Margi-JMP1978,GelfandD-FAA1979},
this leads to a tri-Hamiltonian structure for each nonlinear system in
the coupled AKNS-Kaup-Newell hierarchy
\be u_t=K_n=
J\frac {\delta H_n}{\delta u}= M  \frac {\delta H_{n-1}}{\delta u}
=N\frac {\delta H_{n-2}}{\delta u} ,\  n \ge 2.
\label{tri-HamiltonianstructureofAKNSKNhierarchy}
\ee
The existence of 
all Hamiltonian functions $H_n$ to satisfy $\frac {\delta H_n}{\delta u}
=\Psi ^n \frac {\delta H_0}{\delta u},\,n\ge 0$, is guaranteed by
the hereditary property of the hereditary symmetry operator $\Phi $. 
They are all common conserved densities for the whole AKNS-Kaup-Newell
hierarchy and commute with each other under
three Poisson brackets associated with $J,M$ and $N$.
This is because, for example, we can compute that
\bea &&
\{H_m,H_n\}_J:=
\int<\frac{\delta H_m}{\delta u},J\frac{\delta H_n}{\delta u}>dx
=\int<\frac{\delta H_m}{\delta u},J\Psi\frac{\delta H_{n-1}}{\delta u}>dx
\nonumber \\ && 
=\int<\frac{\delta H_m}{\delta u},\Phi J\frac{\delta H_{n-1}}{\delta u}>dx
=\int<\Psi \frac{\delta H_m}{\delta u},J\frac{\delta H_{n-1}}{\delta u}>dx
\nonumber \\ &&
=\{H_{m+1},H_{n-1}\}_J
=\cdots = \{H_n,H_m\}_J,\ m<n,\ m,n\ge 0. \nonumber \eea 
It gives rise to the commutativity of the conserved densities
$H_n,\, n\ge 0$, by combining  
the skew-symmetric property of the Poisson brackets.
Furthermore we have
\be [K_m,K_n]=J\frac{\delta }{\delta u}\{H_m,H_n\}=0,\ m,n\ge 0, \ee
which implies that each coupled AKNS-Kaup-Newell system 
has infinitely many commuting symmetries. This may also 
be seen from a zero Lie derivative $L_{u_x}\Phi=0$. The
property of $L_{u_x}\Phi=0$ also guarantees that 
 the hereditary
symmetry operator defined by (\ref{hereditaryoperatorAKNSKN})
is a common recursion operator for all systems in the coupled
AKNS-Kaup-Newell hierarchy (\ref{generalhierarchyAKNSKN}).

\section{Concluding remarks}

We have introduced a set of Hamiltonian operators and presented
some corresponding hereditary symmetry operators.
Therefore a coupled AKNS-Kaup-Newell hierarchy of systems
of soliton equations of real form
is proposed. Zero curvature representations
and tri-Hamiltonian structures
are established for all systems in the hierarchy.

Interestingly this coupled AKNS-Kaup-Newell hierarchy
contains two different reductions of the AKNS hierarchy and
the Kaup-Newell hierarchy.
A natural problem we want to ask
is what conditions could be found for the existence of
similar coupled soliton hierarchies
associated with two or more given soliton hierarchies
and how one constructs such coupled soliton hierarchies
if they exist.

Because our coupled AKNS-Kaup-Newell hierarchy includes
the AKNS hierarchy and the Kaup-Newell hierarchy as two simple reductions,
tri-Hamiltonian structures can be constructed for
the AKNS hierarchy and the Kaup-Newell hierarchy, based on the 
obtained tri-Hamiltonian
structures of the coupled AKNS-Kaup-Newell hierarchy.
The corresponding tri-Hamiltonian structure for
the Kaup-Newell system of nonlinear derivative Schr\"odinger
equations has been raised recently in \cite{MaZ-JPA1998}
and a nonlinearization problem has been manipulated
for the associated spectral problem \cite{ZhouM-LAM1998}.
We believe that some other nice properties
may also be achieved for the
the coupled AKNS-Kaup-Newell hierarchy.

It is worthy pointing out that by using a similar deduction to one in Section
\ref{ZCRofcoupledAKNSKN},
a general hereditary symmetry operator defined by
(\ref{hereditaryoperator3}) can be constructed from
the following spectral problem
\be  \phi _x=U\phi,\ U=U(u,\la )=
\left (\ba {cc} \frac 1 {2\gamma }\la &q  \vspace{2mm}\\ 
\frac 1 {2\gamma }(\al -\frac {\beta }{\gamma }\la )r& -\frac 1 {2\gamma }\la 
\ea \right) \ee
with the same constants $\al ,\beta , \gamma $ as ones in
(\ref{hereditaryoperator3}).
It is apparent that
the condition of $\gamma \ne 0$ is required, but $\al $ and
$\beta $ may be equal to zero.
Only a condition
of $\al ^2+\beta ^2\ne 0$ is needed for $\al $ and $\beta $, in order
to guarantee the injective property
of the Gateaux derivative $U'$.
It also deserves to mention that
the Hamiltonian operators defined by (\ref{bigoperator})
can lead to other hierarchies of systems of evolution equations.
For example, a hierarchy of bi-Hamiltonian systems 
$u_t=\Phi ^nu_x,\,n\ge 0$,
can be generated from a hereditary symmetry
operator $\Phi $ defined by (\ref{specialrecursionoperatorAKNS}).
What is more,
we can make another choice of an invertible Hamiltonian operator
\be
 {J} = \left (\ba {cc} 0&\part  \vspace{2mm}\\ \part  &0 
 \ea  \right),\ee
which has an inverse operator
\[
 {J^{-1}} = \left (\ba {cc} 0&\part ^{-1} \vspace{2mm}\\ \part ^{-1} &0 
 \ea  \right).
   \]
It constitutes a Hamiltonian pair together 
with $M$ defined by (\ref{bigoperator}). 
Thus we can have the corresponding hereditary symmetry operator
\be
 \Phi  = MJ^{-1}=\left (\ba {cc} \al _2\part ^{-1}
 +\al _3 -\al _1q\part ^{-1}r\part ^{-1}
 &\al _1 q \part ^{-1}q\part ^{-1}  \vspace{2mm}\\
 \al _1r \part ^{-1}r\part ^{-1}  &
 -\al _2\part ^{-1}+\al _3-\al _1r\part ^{-1}q  \part ^{-1}
 \ea  \right).\ee
 This generates a new hierarchy
 $u_t=\Phi ^nu_x,\, n\ge 0$, which is
 an inverse hierarchy of the 
 Kaup-Newell hierarchy.
 In conclusion, Hamiltonian operators of the same type
may lead to soliton hierarchies of different types. 

\vskip 3mm
\noindent{\bf Acknowledgments:} This work was  supported by
the City University of Hong Kong, the  
Research Grants Council of Hong Kong,
the Chinese National Basic Research Project 'Nonlinear Science', and the
Doctoral Programme Foundation of Ministry of Higher Education, China.


\newpage
\small 
\baselineskip 13pt
\begin{thebibliography}{99}
\bibitem{Newell-book1985}A. C. Newell, {\it Solitons in mathematics and 
Physics} (SIAM, Philadelphia, 1985).
\bibitem{AblowitzC-book1991} M. J. Ablowitz and P. A. Clarkson,
  {\it Solitons, Nonlinear Evolution Equations and Inverse Scattering}
  (Cambridge University Press, Cambridge, 1991).
\bibitem{FokasF-PLA1981}A. S. Fokas and B. Fuchssteiner,
  Phys. Lett. A {\bf 86}, 341--345 (1981).
\bibitem{Tu-SCA1989}G. Z. Tu, Sci. in China A {\bf 32}, 142--153 (1989). 
\bibitem{Ma-CAMB1997} W. X. Ma, Chin. Ann. Math. B {\bf 18}, 79--88 (1997).
\bibitem{BoitiPC-NCB1983}M. Boiti, P. J. Caudrey and F. Pempinelli,
  Nuovo Cimento B {\bf 83}, 71--87 (1984).
\bibitem{AntonowiczF-PD1987}M. Antonowicz and A. P. Fordy,
 Physica D {\bf 28}, 345--357 (1987).
\bibitem{AntonowiczF-JPA1988}M. Antonowicz and A. P. Fordy,
 J. Phys. A: Math. Gen.  {\bf 21}, L269-L275 (1988).
\bibitem{Ma-JPA1993}W. X. Ma, J. Phys. A: Math. Gen.
  {\bf 26} L1169--L1174; W. X. Ma and Z. X. Zhou,
   Prog. Theoret. Phys. {\bf 96}, 449--457 (1996).
\bibitem{MaF-PLA1996}W. X. Ma and B. Fuchssteiner,
  Phys. Lett. A {\bf 213}, 49--55 (1996).
\bibitem{Zhang-ADE1997}B. C. Zhang, Ann. of Diff. Eqs.
  {\bf 13}, 408--418 (1997).
\bibitem{Magri-book1980}F. Magri,  
  in: {\it Nonlinear Evolution Equations and Dynamical Systems},
  Lectures Notes in Physics Vol. 120, eds. M. Boiti, F. Pempinelli and
  G. Soliani (Springer-Verlag,
  Berlin), pp233--263(1980).
\bibitem{Ma-1998}W. X. Ma, 
J. Phys. A: Math. Gen. {\bf 31}, 7585--7591 (1998).
\bibitem{MaP-1998}W. X. Ma and M. Pavlov,
Phys. Lett. A {\bf 246}, 511--522 (1998).
\bibitem{FuchssteinerF-PD1981} B. Fuchssteiner and A. S. Fokas,
  Physica D {\bf 4}, 47--66 (1981).
\bibitem{MaS-PLA1994} W. X. Ma and W. Strampp,
  Phys. Lett. A {\bf 185}, 277-286 (1994).
\bibitem{MaDZL-NCB1996}W. X. Ma, Q. Ding, W. G. Zhang and B. Q. Lu,
  Il Nuovo Cimento B {\bf 111}, 1135--1149 (1996).
\bibitem{Margi-JMP1978}F. Margi, J. Math. Phys. {\bf 19}, 1156--1162 (1978).
\bibitem{GelfandD-FAA1979}I. M. Gel'fand and I. Y. Dorfman,
  Funct. Anal. Appl. {\bf 13}, 248--262 (1979).
\bibitem{MaZ-JPA1998} W. X. Ma and R. G. Zhou, preprint (1998).
\bibitem{ZhouM-LAM1998} R. G. Zhou and W. X. Ma, preprint (1998).

\end{thebibliography}
\end{document}